# Absence of evidence of spin transport through amorphous $Y_3Fe_5O_{12}$

Juan M. Gomez-Perez[a], Koichi Oyanagi[b], Reimei Yahiro[b], Rafael Ramos[c], Luis E. Hueso[a,d], Eiji Saitoh[b,c,e,f,g] and Fèlix Casanova[a,d,*]

[a]CIC nanoGUNE, 20018 Donostia-San Sebastian, Basque Country, Spain
[b]Institute for Materials Research, Tohoku University, Sendai 980-8577, Japan
[c]WPI-Advanced Institute for Materials Research, Tohoku University, Sendai 980-8577, Japan
[d]IKERBASQUE, Basque Foundation for Science, 48013 Bilbao, Basque Country, Spain
[e]Center for Spintronics Research Network, Tohoku University, Sendai 980-8577, Japan
[f]Advanced Science Research, Japan Atomic Energy Agency, Tokay 319-1195, Japan
[g] (Present address) Department of Applied Physics, The University of Tokyo, Tokyo 113-8656, Japan

[*]Email: f.casanova@nanogune.eu

## Abstract

Long-distance transport of spin information in insulators without long-range magnetic order has been recently reported. Here, we perform a complete characterization of amorphous $Y_3Fe_5O_{12}$ (a-YIG) films grown on top of $SiO_2$. We confirm a clear amorphous structure and paramagnetic behavior of our a-YIG films, with semiconducting behavior resistivity that strongly decays with increasing temperature. The non-local transport measurements show a signal which is not compatible with spin transport and can be attributed to the drop of the a-YIG resistivity caused by Joule heating. Our results emphasize that exploring spin transport in amorphous materials requires careful procedures in order to exclude the charge contribution from the spin transport signals.

Insulator-based spintronics is attracting a great amount of attention for the storage and transport of spin information because of the long spin propagation lengths [1] and the absence of energy dissipation due to ohmic losses [2] when compared to the conventional metal-based spintronics. In a ferri- and antiferromagnetic insulator, spin waves or magnons can carry spin information [1,2,11–15,3–10]. In particular, yttrium iron garnet ($Y_3Fe_5O_{12}$, YIG) is a ferrimagnetic insulator which has been broadly studied because of its small damping constant [16] and magnon propagation up to tens of microns [4] or even few millimeters [1,3].

Up to now, long distance spin transport has been achieved in high-quality single-crystal magnetic insulators [1,2,11–15,17,18,3–10]. Moreover, recent reports have stimulated insulator-based spintronics by expanding the field to amorphous materials without long-range magnetic order [19,20]. Amorphous materials are promising for future spintronic devices due to the ease for mass production, small magnetic anisotropy, and less influence of the phonons because of the lack of crystal structure. The work of D. Wesenberg *et. al.* highlights the advantage of amorphous materials for long-range transport of spin information through an amorphous paramagnetic $Y_3Fe_5O_{12}$ without any external magnetic field applied [19]. However, in contrast to single crystals, reproducing the composition and structure of amorphous materials is difficult because they highly depend on the fabrication conditions and treatments. Therefore, in order to establish amorphous-based spintronics, the relation between material properties and spin transport efficiency should be elucidated.

Here, we test the long-distance spin transport reported in Ref. [19] by using amorphous YIG (a-YIG) films grown by magnetron-sputtering deposition on $SiO_2$. We first clarify the amorphous structure and chemical composition of our films by X-ray diffraction (XRD) and transmission electron microscopy (TEM)/scanning TEM (STEM). The magnetic properties are studied by vibrating sample magnetometry (VSM), exhibiting a paramagnetic behavior from room temperature down to 70 K, where an asperomagnetic order appears, which is typical for amorphous magnets [21,22]. We study the spin transport with non-local measurements using the same measurement configuration as in Ref. [19] and standard local resistivity measurements. Our results reveal that the non-local voltage signal observed at room temperature arises from leakage current through the a-YIG film, caused by the rapidly decreasing resistivity of the a-YIG with increasing temperature and, thus, cannot be attributed to spin transport.

Amorphous YIG films were grown on a $SiO_2$ (100 nm)/Si (001) substrate by RF-magnetron-sputtering from a stochiometric $Y_3Fe_5O_{12}$ target. 200-nm-thick films were deposited at 150 W RF-power in an Ar pressure of 0.75 mTorr with a growing rate of 0.063Å/s at room temperature. No post-annealing was performed to the grown films to avoid YIG recrystallization [23]. The characterization of the a-YIG structure and magnetic properties were carried out by TEM/STEM and VSM, respectively. XRD measurements were performed in an equipment with a two-dimensional detector. TEM/STEM was performed on a Titan 60-300 electron microscope (FEI Co., The Netherlands) equipped with EDAX detector (Ametek Inc., USA), high angular annular dark field (HAADF)-STEM detector and imaging Cs corrector. High resolution TEM (HR-TEM) images were obtained at 300 kV at negative Cs imaging conditions [24] so that atoms look bright. Composition profiles were acquired in STEM mode with drift correction utilizing energy dispersive X-ray spectroscopy (EDX) signal. Geometrical Phase Analysis (GPA) was performed on HR-TEM images using all strong reflections for noise suppression [25]. VSM measurements were performed in a liquid-He cryostat (with a temperature $T$ between 2 and 300 K, externally applied magnetic field $\mu_0H$ up to 5 T).

Pairs of Pt strips (width 1 μm, length 200 μm) were prepared on top of the a-YIG film. The strips were patterned by positive *e*-beam lithography with 10 μm distance between the Pt strips. A 10-nm-thick Pt film was deposited *ex-situ* via DC magnetron sputtering (20 W; 0.75 mTorr). In addition, a control sample was prepared, consisting of a Pt injector (width 270 nm, length 72 μm) and two non-local detectors of Pt and Cu, one at each side of the injector. The distance between the injector and each detector is 5 μm. The 5-nm-thick Pt (80 W; 3 mTorr) and 35-nm-thick Cu (250 W; 2 mTorr) were also prepared by *ex-situ* DC magnetron sputtering. Transport measurements were performed in a liquid-He cryostat (temperatures ranging from 10 to 400 K) with a superconducting solenoid magnet (magnetic fields up to 9 T in a rotatable sample stage). The non-local voltage $V_{NL}$ is measured by reversing the DC charge current applied in the injector so that the difference corresponds to the electrical response, $V_{NL} = [V_{NL}(+I) - V_{NL}(-I)]/2$ (equivalent to the 1$^{st}$ harmonic voltage measured with an AC lock-in technique) [4].

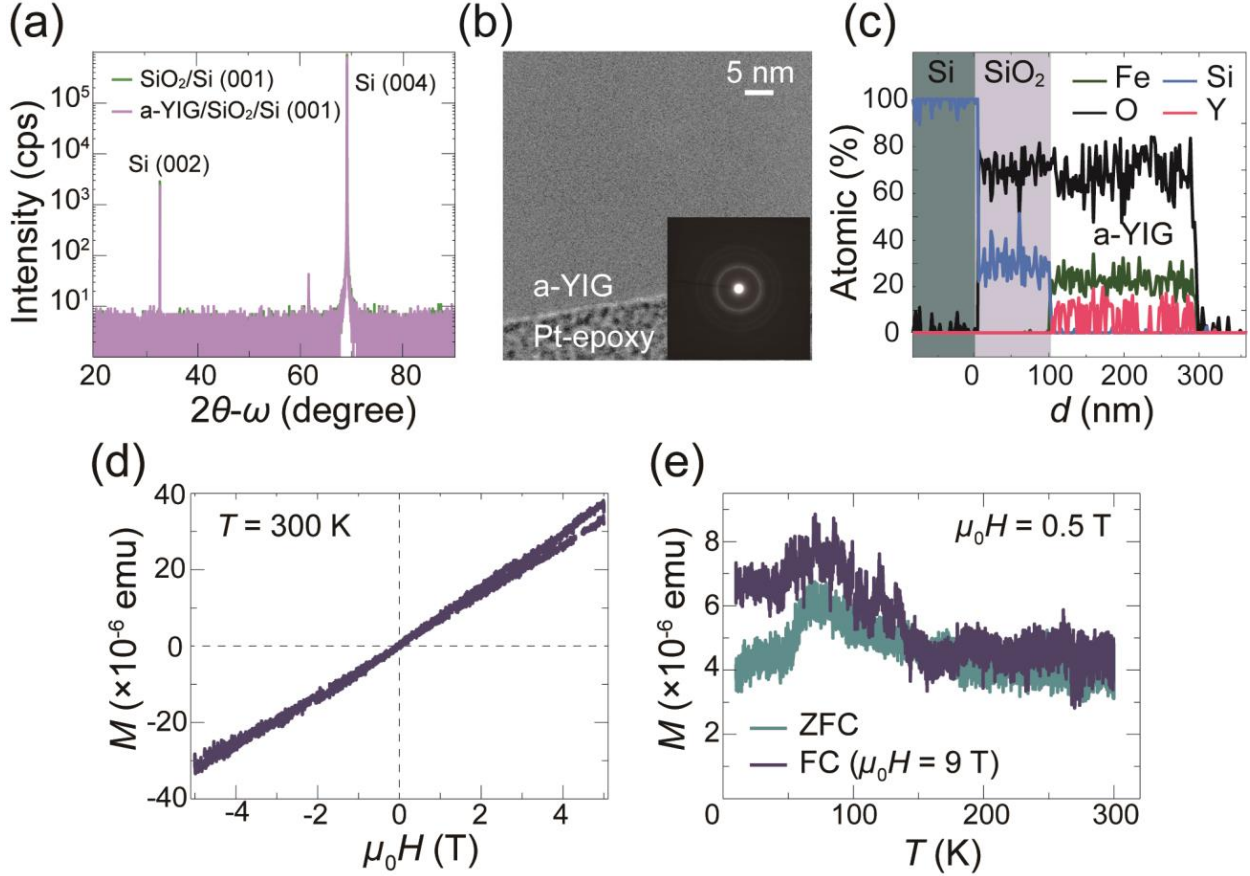

FIG. 1. (a) X-ray diffraction measurement of the SiO$_2$/Si (001) substrate (green) and the a-YIG film growth on top of SiO$_2$/Si (001) (purple). (b) A TEM image of the 200-nm-thick a-YIG with an inset of the diffraction pattern of the film. (c) EDX analysis of spatial distribution of the elements along the out-of-plane direction, where the depth equal to 0 nm corresponds to the interface between SiO$_2$ and Si. The green, black, blue and pink lines correspond to the Fe, O, Si and Y atomic concentration (in %), respectively. (d) Magnetization as a function of applied in-plane magnetic field (trace and retrace) at room temperature of the a-YIG film. The diamagnetic background from the SiO$_2$/Si substrate was subtracted after measuring it from a reference sample. (e) Zero-field-cooled (ZFC)–field-cooled (FC) magnetization of a-YIG as a function of temperature measured on heating at $\mu_0 H = 0.5$ T. The dark blue (light blue) line corresponds to the FC (ZFC) curve after cooling at $\mu_0 H = 9$ T (0 T). The same diamagnetic background from the SiO$_2$/Si substrate was also subtracted and assumed to be temperature independent.

Figure 1(a) shows the XRD spectrum of a-YIG/SiO$_2$/Si (001) sample (purple curve) and a SiO$_2$/Si (001) substrate (green curve). In both cases, the same two peaks can be seen at 32.9° and 69.1° corresponding to Si (002) and Si (004). The absence of the peaks corresponding to crystal YIG [26] in the purple curve suggests the amorphous behavior of the film. We performed TEM bright field (Fig. 1(b)) images that show a uniform amorphous structure. The inset of Fig. 1(b) shows a clear halo pattern of the a-YIG lattice, which is the typical diffraction pattern for amorphous materials. We also performed EDX analysis of the spatial distribution of the elements along the out-of-plane direction (Fig. 1(c)). By extracting the mean and the standard error of the atomic % data along the profile, we found that the concentration along the 200-nm-thick a-YIG for Fe, O and Y is constant within the error bars. The film composition is slightly different compared with stochiometric YIG (Y$_3$Fe$_5$O$_{12}$) [27]. The Fe and Y content is 21.9±0.4% and 12±1%, respectively, being a bit smaller than that of stochiometric YIG (where the Fe and Y content are 25% and 15%, respectively). The film is more oxidized than expected, with an O content of 68±1% instead of the stochiometric 60%.

Figures 1(d) and (e) show the a-YIG magnetization $M$ as a function of the magnetic field $H$ and temperature $T$, respectively. In both cases, the $T$-independent diamagnetic background of the SiO$_2$/Si measured in a reference substrate has been subtracted. The $M(H)$ curve at 300 K shows a linear behavior with no hysteresis between $-5$ T and 5 T, characteristic of the paramagnetic phase. The field-cooled (FC) curve was performed by first cooling the sample with a strong in-plane magnetic field applied (9 T) and then measuring the magnetization $M$ as a function of $T$ while heating under a small in-plane applied magnetic field (0.5 T). For the case of zero-field-cooled (ZFC) curve, the sample was cooled down from room temperature without any applied magnetic field and, after that, a small in-plane magnetic field was applied to measure $M$ while increasing the temperature. The FC curve gradually increases down to 70 K, being consistent with a typical Curie-Weiss trend of paramagnetic materials. However, below 70 K we find a decrease of $M$. This decrease, as well as the gap opening at low temperatures between FC and ZFC curves, are attributed to the phase transition of the a-YIG from a paramagnetic to an asperomagnetic phase [28,29]. In the asperomagnetic phase, random magnetic moments add up to a non-vanishing macroscopic magnetization [30]. Hence, based on the structural and magnetic features, we confirm the amorphous and paramagnetic nature of our a-YIG films.

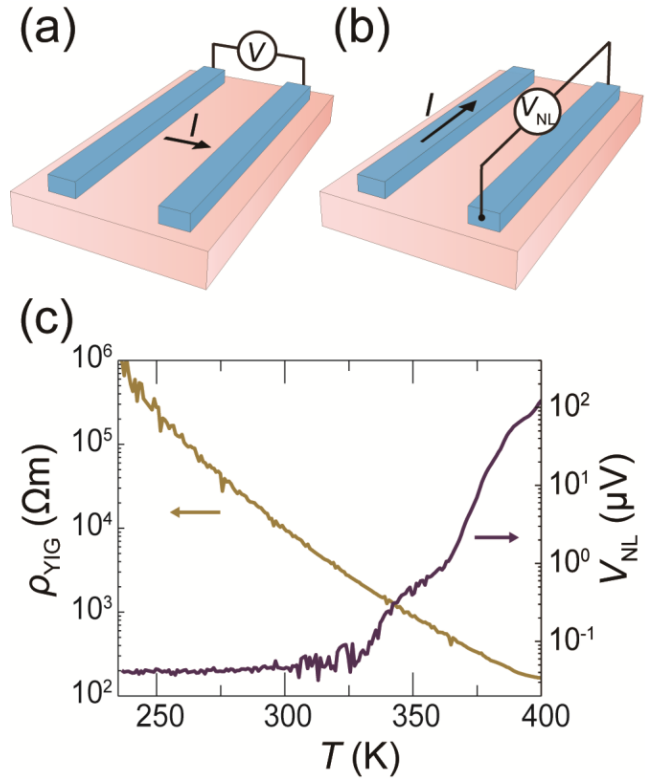

FIG. 2. Schematics of measurement configuration for (a) the YIG resistivity $\rho_{\text{YIG}}$ and (b) the non-local voltage $V_{\text{NL}}$ across the Pt strip. (c) Semi-log plot of the temperature dependence of $\rho_{\text{YIG}}$ (brown) and $V_{\text{NL}}$ (purple).

We evaluate the electrical properties of the a-YIG film using the configuration sketched in Fig. 2(a). Firstly, we applied a constant voltage of 10 V between the Pt contacts and detected the current flowing through the a-YIG to measure the a-YIG resistivity ($\rho_{\text{YIG}}$). The purple curve in Fig. 2(c) corresponds to the $T$ dependence of $\rho_{\text{YIG}}$, which shows a semiconducting behavior and drops with increasing $T$ from $10^6$ $\Omega\cdot$m at 235 K to $10^2$ $\Omega\cdot$m at 400 K. A similar resistivity drop is reported for ultrathin YIG films [31]. At room temperature, our $\rho_{\text{YIG}}$ is $10^4$ $\Omega\cdot$m, which is six orders of magnitude smaller than that of single-crystal YIG at the same temperature, $\sim 10^{10}$ $\Omega\cdot$m [32].

Next, we performed non-local transport measurements in the very same device (configuration sketched in Fig. 2(b)). As discussed above, a non-local spin transport has been reported in a-YIG and attributed to the correlation-mediated spin current [19]. When a charge current is applied along the Pt wire, a transverse pure spin current is created due to the spin Hall effect (SHE) [33]. This pure spin current generates a spin accumulation at the top and the bottom of the Pt wire. The interfacial exchange interaction would transfer spin angular momentum from Pt electrons to the local $Fe^{3+}$ moments in a-YIG [34] and create a non-equilibrium spin accumulation in a-YIG. If a-YIG could transport this spin accumulation, that would be detected in a second Pt strip *via* the inverse SHE (ISHE) [33,35]. We applied a DC charge current of 3 mA (current density $j_c \sim 10^{11}$ A/m$^2$) along the Pt injector and detected a non-local voltage $V_{\text{NL}}$ across the Pt detector. Figure 2(c) shows the $T$ dependence of $V_{\text{NL}}$ (brown curve). Below 330 K, no detectable signal appears above the background noise of $\sim 2$–$5 \times 10^{-2}$ $\mu$V. However, above 330 K, a non-local signal appears, and its amplitude increases strongly up to $10^2$ $\mu$V at 400 K. Coincidently, the non-local signal increases while $\rho_{\text{YIG}}$ drops by two orders of magnitude in the range between 300 and 400 K. Interestingly, the onset temperature ($T_{\text{onset}}$) of $\sim 330$ K is compatible with the one reported in Ref. [19].

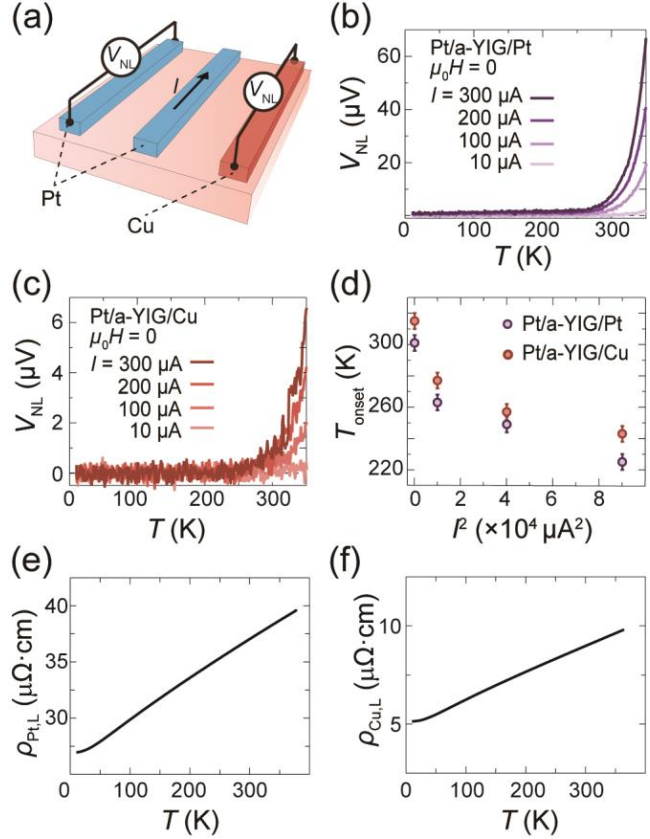

FIG. 3. (a) Non-local measurement configuration used in the control sample, with a Pt injector and two non-local detectors, Pt (left) and Cu (right). (b-c) Temperature dependence of the non-local voltage $V_{NL}$ detected at (b) the Pt detector and (c) the Cu detector for different charge currents applied. (d) Onset temperature $T_{onset}$ as a function of the square of the applied current for the Pt and the Cu detectors. (e-f) Temperature dependence of the (e) Pt resistivity and (f) Cu resistivity.

To understand the origin of $V_{NL}$, we performed non-local measurements in the control sample (see Fig.3(a)). The middle Pt strip is used as an injector and the other two are used as non-local detectors. One of them is made of Pt and the other one is made of Cu, which has a weak spin-orbit coupling and thus shows negligible SHE. We expect that, if the ISHE governs the $V_{NL}$ detected above $T_{onset}$, $V_{NL}$ will disappear for the Pt/a-YIG/Cu configuration. Figures 3(b) and (c) show the $T$ dependence of $V_{NL}$ at different currents $I$ for the Pt and Cu detectors, respectively. For the Pt detector, we observed the very same trend shown in Fig. 2(c); $V_{NL}$ appears above certain $T_{onset}$ and increases with $T$. However, in contrast to our expectation, we also measured a clear $V_{NL}$ across the Cu detector with the same trend as the Pt case, indicating that the observed $V_{NL}$ cannot be attributed to the spin current flowing through the a-YIG film, which is actually supported by a theoretical work [36], where they have found that the spin angular momentum cannot transfer through such disordered system.

Figure 3(d) shows $T_{onset}$ in the Pt and Cu detectors as a function of $I^2$. Both $T_{onset}$ show the same trend; the signal appears at lower $T$ with larger $I^2$. These results can be easily understood by considering that the device temperature increases due to Joule heating, resulting in the decrease of $\rho_{YIG}$ (the semiconducting trend shown in Fig. 2(c)). Consequently, $V_{NL}$ in both Pt and Cu detectors arise from the charge transport through the a-YIG film, because the nonlocal configuration we use is consistent with the van der Pauw configuration [37], and could also explain the results in Ref. [19]. This interpretation also explains why the amplitude of $V_{NL}$ in the Cu detector is one order smaller than in the Pt detector: the higher resistivity of Pt as compared to Cu gives rise to a higher detected voltage for the same flowing current, see Figs. 3(e) and 3(f). The difference between the $T_{onset}$ for the Cu and Pt detectors is related to this amplitude difference of the $V_{NL}$, the lower signal-to-noise ratio in Cu makes the extraction of an accurate value of $T_{onset}$ difficult. Finally, if we compare our results to the ones published in Ref. [19], we can evidence the similarities in the temperature dependence of $V_{NL}$, as well as the similar $T_{onset}$ at which we are able to detect the non-local voltage. In our case, though, we clearly demonstrate the leakage current between the contacts due to the local raise of temperature as the origin of the effect.

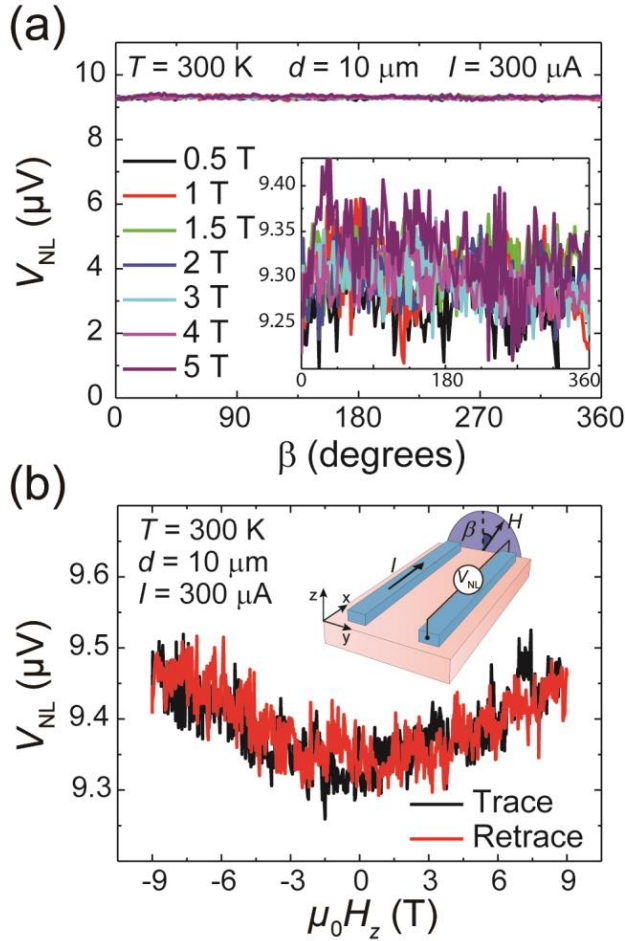

Fig. 4. (a) Non-local ADMR measurements in the $zy$−plane, where the magnetic field is rotated an angle $\beta$ from the out-of-plane to an in-plane direction (always perpendicular to the applied current, see sketch in panel b), at different magnetic fields and $T = 300$ K. Inset: zoom of the measurements. (b) Non-local FDMR measurements with the magnetic field applied in the out-of-plane $z$−direction, at $T = 300$ K. Inset: sketch of the measurement configuration.

In addition, we performed non-local angular dependent magnetoresistance (ADMR) measurements to confirm that $V_{NL}$ is not related to spin or magnon transport. Figure 4(a) shows the ADMR with the magnetic field rotating an angle $\beta$ in the $zy$−plane [see sketch in Fig. 4(b) for the measurement configuration] at 300 K and different fields up to 5 T. There is no modulation with $\beta$ as one would expect if $V_{NL}$ is related to spin or magnon transport through a magnetic insulator [4,5]. Moreover, we performed non-local field dependent magnetoresistance (FDMR) measurements with the magnetic field swept in the $z-$direction (out of plane) at 300 K, shown in Fig. 4(b). Only a small positive magnetoresistance with no hysteresis is detected up to 9 T, ruling out again any magnon transport origin of $V_{NL}$, since such a large magnetic field would suppress magnons and, thus, $V_{NL}$ [38]. Considering that the origin of $V_{NL}$ is the current leakage through the film, the observed FDMR is most likely ordinary magnetoresistance of the a-YIG film, since no anisotropic magnetoresistance is expected to be present in the a-YIG at this temperature, which is paramagnetic.

In conclusion, we carefully studied the spin transport in amorphous sputtered $Y_3Fe_5O_{12}$ films, as previously reported in Ref. [19]. The magnetic characterization shows a paramagnetic behavior from room temperature down to 70 K and, below 70 K, the asperomagnetism previously reported in a-YIG. Through the local and non-local transport measurements at different temperatures, applied currents and magnetic fields, we conclude that the observed non-local voltage signal does not correspond to spin or magnon transport and can only be attributed to a leakage current between the injector and detector due to the dramatic drop of the a-YIG resistivity induced by local Joule heating.

**Acknowledgements**

The authors thank Andrey Chuvilin for TEM/STEM images and the compositional analysis performed in CIC nanoGUNE. We also thank S. Ito from the Institute for Materials Research, Tohoku University, for performing TEM and STEM on our samples. The work was supported by the Spanish MINECO under the Maria de Maeztu Units of Excellence Programme (MDM-2016-0618) and under Project Nos. MAT2015-65159-R and RTI2018-094861-B-100, the Regional Council of Gipuzkoa (Project No. 100/16), ERATO "Spin Quantum Rectification Project" (No. JPMJER1402) from JST, the Grant-in-Aid for Scientific Research on Innovative Area "Nano Spin Conversion Science" (No. JP26103005), JSPS Core-to-Core program "the International Research Center for New-Concept Spintronics Devices", and World Premier International Research Center Initiative (WPI) from MEXT, Japan. J.M.G.-P. thanks the Spanish MINECO for a Ph.D. fellowship (Grant No. BES-2016-077301), K.O. acknowledges support from GP-Spin at Tohoku University.